\documentclass{optica-article}

\journal{opticajournal} 
\articletype{Research Article}

\usepackage{lineno} 

\AtBeginDocument{\RenewCommandCopy\qty\SI} 

\usepackage{amsmath, amssymb, physics, siunitx}
\usepackage{caption, nicefrac, esdiff}
\usepackage{subcaption}
\usepackage[normalem]{ulem}
\usepackage{booktabs}
\usepackage{hyperref}

\begin{document}

\title{Optimizing the quantum interference between single photons and local oscillator with photon correlations}

\author{Hubert Lam\authormark{1${\dagger}$}, Juan R. Álvarez\authormark{1*}, Petr Steindl\authormark{1,2}, Ilse Maillette de Buy Wenniger\authormark{3}, Stephen Wein\authormark{4}, Anton Pishchagin\authormark{4}, Thi Huong Au\authormark{4}, Sebastien Boissier\authormark{4}, Aristide Lemaître\authormark{1}, Wolfgang Löffler\authormark{2}, Nadia Belabas\authormark{1}, Dario A. Fioretto\authormark{1,4}, and Pascale Senellart\authormark{1}}

\address{
\authormark{1}C2N,~Photonics~Department,~Université~Paris-Saclay,~10~Bd~Thomas~Gobert,~91120~Palaiseau,~France \\
\authormark{2}Leiden Institute of Physics, Leiden University, 2333 CA Leiden, The Netherlands \\
\authormark{3}Imperial College London, Quantum Optics and Laser Science, Exhibition Rd, London SW7 2BX, UK
\authormark{4}Quandela, 7 Rue Léonard de Vinci, 91300 Massy, France\\
}

\email{\authormark{$\dagger$}hubert.lam@universite-paris-saclay.fr} 

\address{{\scriptsize \authormark{*}Currently at LCTI,~Telecom~Paris,~IP~Paris,~19~Place~Marguerite~Perey,~91120~Palaiseau,~France.}}

\begin{abstract*} 
The quantum interference between a coherent state and a single photon is an important tool in continuous variable optical quantum technologies to characterize and engineer non-Gaussian quantum states. Semiconductor quantum dots, which have recently emerged as a key platform for efficient single-photon generation, could become interesting assets in this context. An essential parameter for interfering single photons and classical fields is the mean wavepacket overlap between both fields. Here, we report on two homodyne photon-correlation techniques enabling the precise measurement of the overlap between a single photon generated by a quantum dot-cavity device and pulsed laser light. The different statistics of interfering fields lead to specific signatures of the quantum interference on the photon correlations at the output of the interfering beam splitter. We compare the behavior of maximized overlap, measuring either the Hong-Ou-Mandel visibility between both outputs or the photon bunching at a single output. Through careful tailoring of the laser light in various degrees of freedom, we maximize the overlap to $76\,\%$, with limitations primarily due to mismatched spectral and temporal profiles and low-frequency charge noise in the single-photon source.
\end{abstract*}

\newpage
\section{Introduction}
Exploiting light to process quantum information is a continuously growing field, with foreseen applications in quantum computing~\cite{maring2024versatile, aghaee_rad_scaling_2025}, quantum communications~\cite{chen_integrated_2021, chen_twin-field_2021, hajomer_long-distance_2024}, and quantum sensing~\cite{jia_squeezing_2024}. Single-photon states are central to these technological developments, both when encoding the information on single photons (discrete variable framework) or in the quadrature of the quantum light field (continuous variable framework). Single photons have been extensively exploited to implement discrete variable quantum information processing protocols~\cite{maring2024versatile}. Having also intrinsically non-Gaussian character, single photons can be used to implement non-Gaussian operation on a Gaussian state -- a necessary gate for universal continuous variable quantum computing~\cite{Lloyd1999}. Such gate can be realized by a heralded interference of the single photon and the Gaussian state, effectively leading to single-photon addition~\cite{Dakna1998} and generation of a non-Gaussian state~\cite{Dakna1997, Barnett2018, Zavatta2007}.

Concretely, interfering a single photon and a coherent state on a beam splitter is a powerful tool in this framework. For instance, such interference combined with a proper heralding scheme allows the generation of a variety of non-Gaussian states, including displaced Fock states and photon-number superpositions~\cite{Lvovsky2002_photoncatalysis, Lvovsky2002, eaton_non-gaussian_2019}. Finally, note that the non-Gaussianity of a single-photon state in the continuous variable framework is assessed using homodyne detection through the interference with a laser light acting as local oscillator~\cite{Paris1996, lvovsky_continuous-variable_2009}.

So far, the vast majority of experiments interfering single photons with coherent states have been implemented with heralded single-photon sources based on frequency conversion ~\cite{laiho2009producing,Koashi1996,Laiho2012,Shen2017}. These sources  are based on a probabilistic generation process and suffer from intrinsic limitation on terms of efficiency. In the last decade, a new technology to generate  single photons in a deterministic way has emerged based on semiconductor quantum dots (QDs)~\cite{Somaschi2016, Tomm2021, Thomas2021, Ding2016, ding_high-efficiency_2025}. Exploiting these sources  would bring the advantage of on-demand deterministic generation with integrated devices to
the continuous variable framework.

In the present work, we perform a first step in this direction and quantify the interference between a laser pulse and the single photons generated by a QD in a cavity~\cite{Somaschi2016}. We report on two complementary experimental methods to assess and optimize their mode overlap $M$ (called mean wavepacket overlap)~\cite{Vogel1995}. The first is reminiscent of the Hong-Ou-Mandel (HOM) interference performed with two single photons, where $M$ is directly related to the interference visibility determined by cross-correlation measurements between the two outputs of the beam splitter~\cite{Hong1987, ollivier2021hong}. The second method monitors the photon bunching through two-photon correlation at a single output of the beam splitter. Here again, the difference in photon-number statistics between both interfering fields leads to a distinctive signature for maximal overlap, which is obtained by adjusting the power of the laser light.

\begin{figure}[!ht]
         \centering
    \includegraphics[width=\columnwidth]{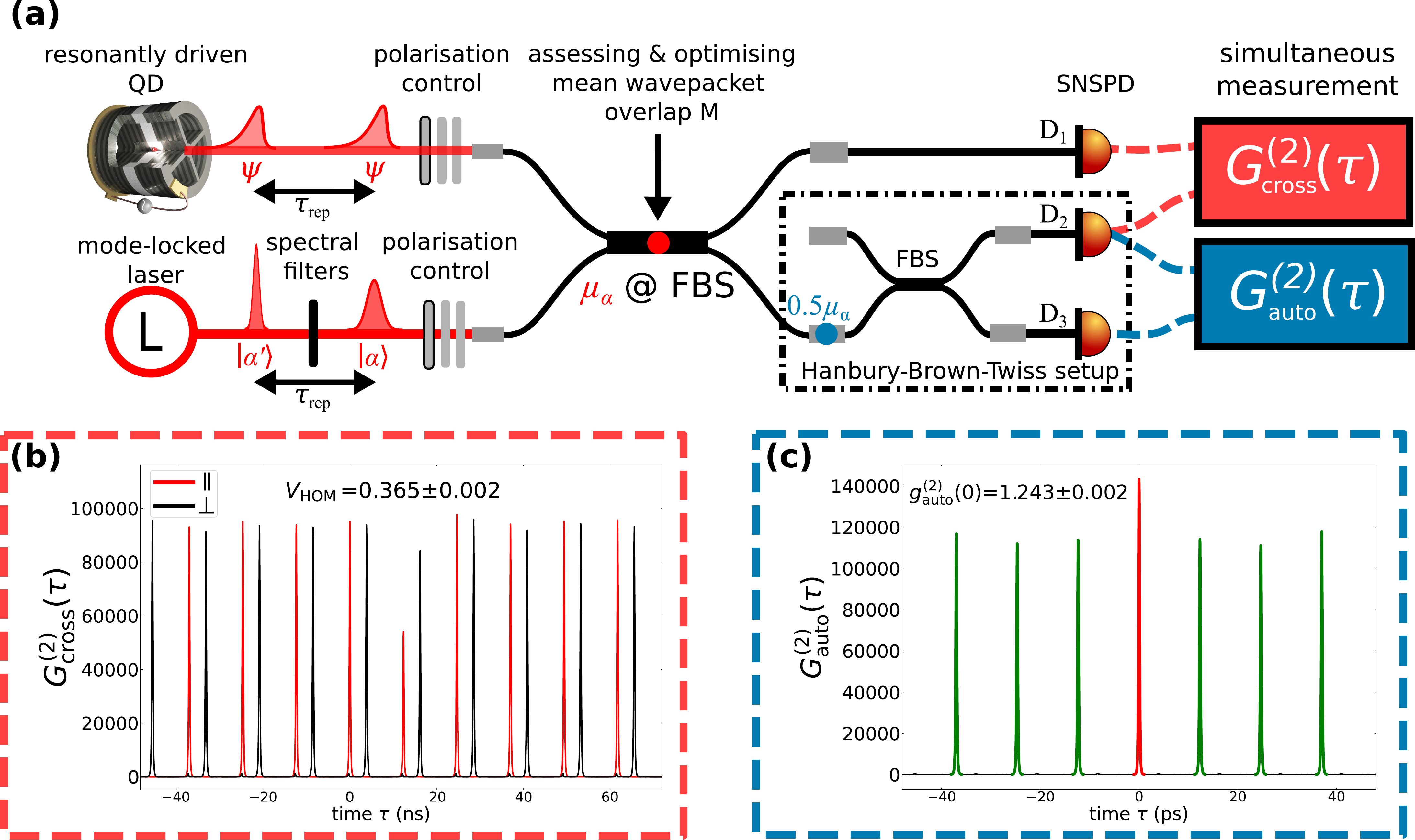}
 \caption{
(a) Experimental setup: weak laser light pulses $\ket{\alpha}$ interfere at a balanced FBS with single photons $\psi$ emitted by InGaAs QD in a micropillar cavity after coherent and resonant excitation with a mode-locked Ti:Sapphire laser. The coherent state $\ket{\alpha}$ originates from the same laser. 
Both FBS outputs are analyzed simultaneously in cross- and auto-correlation configurations on SNSPD pairs ($D_1$, $D_2$) and ($D_2$, $D_3$), respectively. 
(b) Cross-correlation histogram shows reduced coincidences in the presence of HOM interference (parallel $\parallel$, red) compared to the absence of the interference (orthogonal $\perp$, black). (c) Auto-correlation histogram shows bunching statistics.
} \label{fig:g2_cross_auto_sketch_histo}
\end{figure}   

\section{Experimental procedure}
Our single-photon sources are based on InGaAs QDs grown by molecular beam epitaxy embedded in the center of a $\lambda$ GaAs optical cavity between two AlGaAs/GaAs distributed Bragg reflectors with 18 top and 36 bottom pairs. Furthermore, a single QD is deterministically positioned at the electric field antinode of the subsequently etched micropillar cavity with approximately $50\,$nm accuracy using the in-situ lithography technique~\cite{Dousse2008}. The QD is embedded in a p-i-n junction, allowing tuning of its optical transition to cavity resonance around $\sim\!925\,$nm by the quantum-confined Stark effect, leading to a Purcell-enhanced radiative lifetime $\tau_\mathrm{LT}$ of $(170\pm5)\,$ps. The QD-cavity system operates at $5\,$K in a closed-cycle cryostat. We create single photons by coherently exciting the QD micropillar system on-resonance using a mode-locked Ti:Sapphire laser with $\sim\!10\,$ps pulse duration and $1/\tau_\mathrm{rep}\approx82\,$MHz repetition rate. The maximum population inversion of the QD excitonic states is achieved with a pulse area corresponding to a $\pi$. This results in a normalized second-order intensity correlation of $g^{(2)}_{\psi}(0)=4.12\pm0.05\,\%$ where $\psi$ indicates the averaged wavepacket emitted by the QD device. The indistinguishability of successively emitted photons is measured to reach $M_\psi=(90.5\pm0.1)\,\%$~\cite{ollivier2021hong}.

The emitted photonic state is separated from the excitation laser by polarization filtering in a cross-polarization setup~\cite{Somaschi2016, Steindl2023_PER}. The collected photons are directed to a single-mode fiber-based beam splitter (FBS) of near-balanced splitting ratio, ensuring near-ideal spatial overlap, where the photons are mixed with a power-controlled coherent state $\ket{\alpha}$ supplied from the same excitation laser. The coherent state is spectrally shaped to ensure mode overlap in the spectral-temporal domain with the single photons, as illustrated in Fig.~\ref{fig:g2_cross_auto_sketch_histo}(a). The spectral shaping of the coherent state is performed by a pair of Fabry-Pérot etalons ($\mathrm{FWHM}= $14.5\,pm) to approach the spectral-temporal properties of $\psi$ - details in the Supplementary Sec.~\ref{subsec:MeanWavePacketOverlap_independent}. The power of the local oscillator is controlled using two calibrated variable optical attenuators. We control the polarization of each field with a separate polarizer followed by a set of free-space zero-order waveplates. The timing of laser and single photons arrival is adjusted finely using free-space delay lines.

We detect light at both FBS outputs using superconducting nanowire single-photon detectors (SNSPDs) and simultaneously measure cross- and auto-correlations \cite{Vogel1995} between the FBS outputs using three detectors, as illustrated in Fig.~\ref{fig:g2_cross_auto_sketch_histo}(a). Examples of the correlations $G^{(2)}(\tau)$ measured in cross-correlation and auto-correlation configurations with a local oscillator power of $\mu_\alpha=0.28\pm0.01$ are shown in Figs.~\ref{fig:g2_cross_auto_sketch_histo}(b) and \ref{fig:g2_cross_auto_sketch_histo}(c). 

\section{HOM visibility analysis} 
First, we consider the cross-correlation measurement where the overlap $M$ between two optical modes can be deduced from the strength of photon bunching on a balanced beam splitter. This method is commonly used to determine the indistinguishability of photons successively emitted by a QD source, where the total mean wavepacket overlap $M_\psi = V_\mathrm{HOM}\left(1+g^{(2)}_\psi(0)\right)$~\cite{ollivier2021hong} can be quantified by the HOM visibility $V_\mathrm{HOM}$ and the purity of the single-photon stream $g^{(2)}_\psi(0)$. This approach corrects for multi-photon contamination in the QD emission, which reduces the strength of photon bunching~\cite{maillette2023experimental, ollivier2021hong}.

\begin{figure}[!ht]
\centering
         \includegraphics[width=\columnwidth]{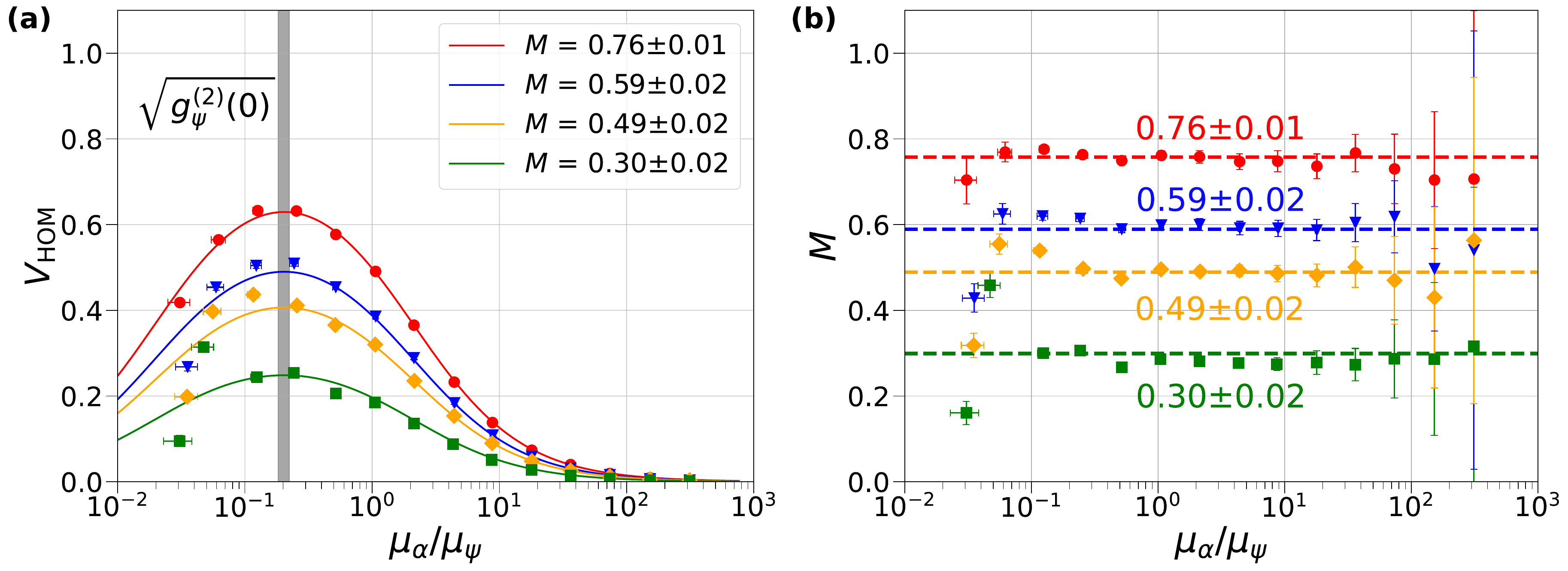}
\caption{
(a)~HOM visibility $V_\mathrm{HOM}$ as a function of relative mean photon number of the fields $\mu_\alpha/\mu_\Psi$. Each color denotes a different $M$ value, which is controlled by altering the relative polarization of the $\ket{\alpha}$-path. The measured $V_\mathrm{HOM}$ is fitted with Eq.~\ref{eq:Mod_HOM_Vis} and the obtained $M$ is given in the legend. Identical $M$ values can be also obtained from the maximum of $V_\mathrm{HOM}$ at $\mu_\alpha/\mu_\psi=\sqrt{g^{(2)}_\psi(0)}$ (gray vertical line). (b)~The overlap $M$ extracted from individual data points match within error bars the values obtained from the fit (a) (horizontal lines).}\label{fig:M_cross}    
\end{figure}

We extend this formalism to the interference between two light fields with dissimilar photon-number statistics: a stream of photons $\psi$ emitted by our QD and a coherent state $\ket{\alpha}$. We start by comparing the photon correlations measured with the polarization direction of the coherent state aligned parallel ${g^{(2)}_{\mathrm{cross},\parallel}(\tau)}$ and orthogonal ${g^{(2)}_{\mathrm{cross},\perp}(\tau)}$ with the polarization of $\psi$, as shown in Fig.~\ref{fig:g2_cross_auto_sketch_histo}(b). Similarly to the conventional Hong-Ou-Mandel experiment, the strength of photon coincidences measured at $\tau=0$ is the litmus test of the photon-bunching effect: the single photons in the photon stream $\psi$ bunch with the single-photon component from the local oscillator $|\alpha\rangle$ leading to reduced correlations at $g^{(2)}_{\mathrm{cross},\parallel}(0)$ with respect to $g^{(2)}_{\mathrm{cross},\perp}(0)$. The relative weight of the single-photon component in the coherent state during interference is inherently constrained by its Poissonian statistics, dictated by the mean photon number $\mu_\alpha=|\alpha|^2$. Consequently, the effective strength of the photon bunching also depends on $\mu_\alpha$. Fig.~\ref{fig:M_cross}(a) presents the HOM visibility ${V_\mathrm{HOM} = ( g^{(2)}_{\mathrm{cross},\perp}(0)-g^{(2)}_{\mathrm{cross},\parallel}(0))/g^{(2)}_{\mathrm{cross},\perp}(0)}$ where the normalized second-order correlations ${g^{(2)}_{\mathrm{cross},\perp}(0)}$ and ${g^{(2)}_{\mathrm{cross},\parallel}(0)}$ are obtained from the measured correlation histograms \cite{bennett2009interference, Li2013, maillette2023experimental,PadronBrito2021}. The experiment is conducted for various polarizations of the interfering fields, from parallel polarization (red symbols) to increasing non-parallel polarizations (blue, yellow, green symbols). For all curves, the HOM visibility approaches zero when $\mu_\alpha$ approaches either zero or large values. Such behavior is expected for both limits, since the strong imbalance of the mean photon numbers between both fields results in an effective single-field interference.

To access the mean wavepacket overlap between both fields, we derive the HOM visibility $V_\mathrm{HOM}$ as a function of the mean wavepacket overlap $M$ and the normalized second-order intensity correlation of each field (see  Supplementary Sec.~\ref{sec:Theory_supp}). The normalized second-order correlation function of the local oscillator $\ket{\alpha}$ is given by $g_\alpha^{(2)}(0)=1$. For QD emission, using a short resonant $\pi$-pulse compared to the emission lifetime, we approximate the emitted state by $p_1|1\rangle\langle1|+p_2|2\rangle\langle2|$,  where $p_1$ and $p_2$ correspond to the probability of the QD emitting one and two photons, respectively. We consider $p_1\approx1$, $p_2\ll p_1$ and neglect any coherence of the photon-number~\cite{loredo_generation_2019}, so that $g^{(2)}_\psi(0)=2p_2/(p_1+2p_2)^2$. The mean wavepacket overlap of both fields is then given by~\cite{maillette2023experimental}:
\begin{equation}
M=V_{\mathrm{HOM}}\left(1+\frac{\mu_\alpha}{{2\mu_\psi}}g_\alpha^{(2)}(0)+\frac{\mu_\psi}{2\mu_\alpha}g_\psi^{(2)}(0)\right)\,,\label{eq:Mod_HOM_Vis}
\end{equation}
where $\mu_\psi=\eta (p_1+2p_2)$ denotes the mean photon number of $\psi$ transmitted through the quantum channel (from QD to FBS) of total transmission $\eta$. The reconstructed $M$ is shown in Fig.~\ref{fig:M_cross}(b) which evidences the maximal mode overlap $M=0.76\pm0.01$ for parallel interfering fields (red). This value is found to be constant for almost four orders of magnitude of $\mu_\alpha$. Such a stability requires a very precise calibration of $\mu_\alpha$ at the beam splitter that accounts for the polarization-dependent detection efficiency of SNSPDs, as explained in Supplementary Sec.~\ref{sec:Calib_mualpha} and~\ref{sec:Suppl_polarizationCor}.

To the best of our knowledge, such a high value of overlap represents the highest value reported with on-demand semiconductor-QD single-photon sources and pulsed laser. However, it remains slightly limited due to off-resonant light leakage through the laser shaping etalons, giving that the overlap increases to $M=0.81\pm0.01$ with post-processing correction (see Supplementary Sec.~\ref{subsec:MeanWavePacketOverlap_independent}). Nonetheless, the corrected overlap achieved in the pulsed regime is nearly equivalent yet remains slightly below the previously reported maximum of 83\,\%~\cite{bennett2009interference} in the continuous-wave regime~\cite{Prtljaga2016,Felle2015}. The remaining distinguishability arises from the mode mismatch between the spectral and temporal profiles of the two fields and the low-frequency charge noise of the single-photon source. These factors lead to a spectral and temporal overlap of $M_f=0.85\pm0.02$ and $M_t=0.910\pm0.005$, respectively (see Supplementary Sec.~\ref{subsec:MeanWavePacketOverlap_independent}).

In addition, Fig.~\ref{fig:M_cross}(b) evidences a control over $M$ for several mutual polarization alignments between the two interfering fields. The relative polarization of the interfering fields is calibrated by monitoring the interference visibility using a narrow-band continuous wave calibration laser (see Supplementary Sec.~\ref{subsec:polar}). For each polarization, we consistently observe the maximum of $V_\mathrm{HOM}=M/{(\sqrt{g^{(2)}_\psi(0)}+1)}$ at $\mu_\alpha/\mu_\psi=\sqrt{g^{(2)}_\psi(0)}=0.203\pm0.001$, as expected \cite{PadronBrito2021}, marked in Fig.~\ref{fig:M_cross}(a) by a gray vertical line. Note that the uncertainties of $M$ in Fig.~\ref{fig:M_cross}(b) for the larger $\mu_\alpha$ become pronounced due to the decrease of $V_\mathrm{HOM}$ to zero and the divergence of the multi-photon correction factor in Eq.~(\ref{eq:Mod_HOM_Vis}). The errors at lower values of $\mu_\alpha$ arise from the birefringence of the power control attenuators at higher voltages, slightly altering the estimation of the mean photon number $\mu_\alpha$ due to the polarizer defining the local oscillator polarization, as discussed in Supplementary Sec.~\ref{sec:Calib_mualpha}.

\section{Bunching analysis}

\begin{figure}[!ht]
    \centering
         \includegraphics[width=0.65\columnwidth]{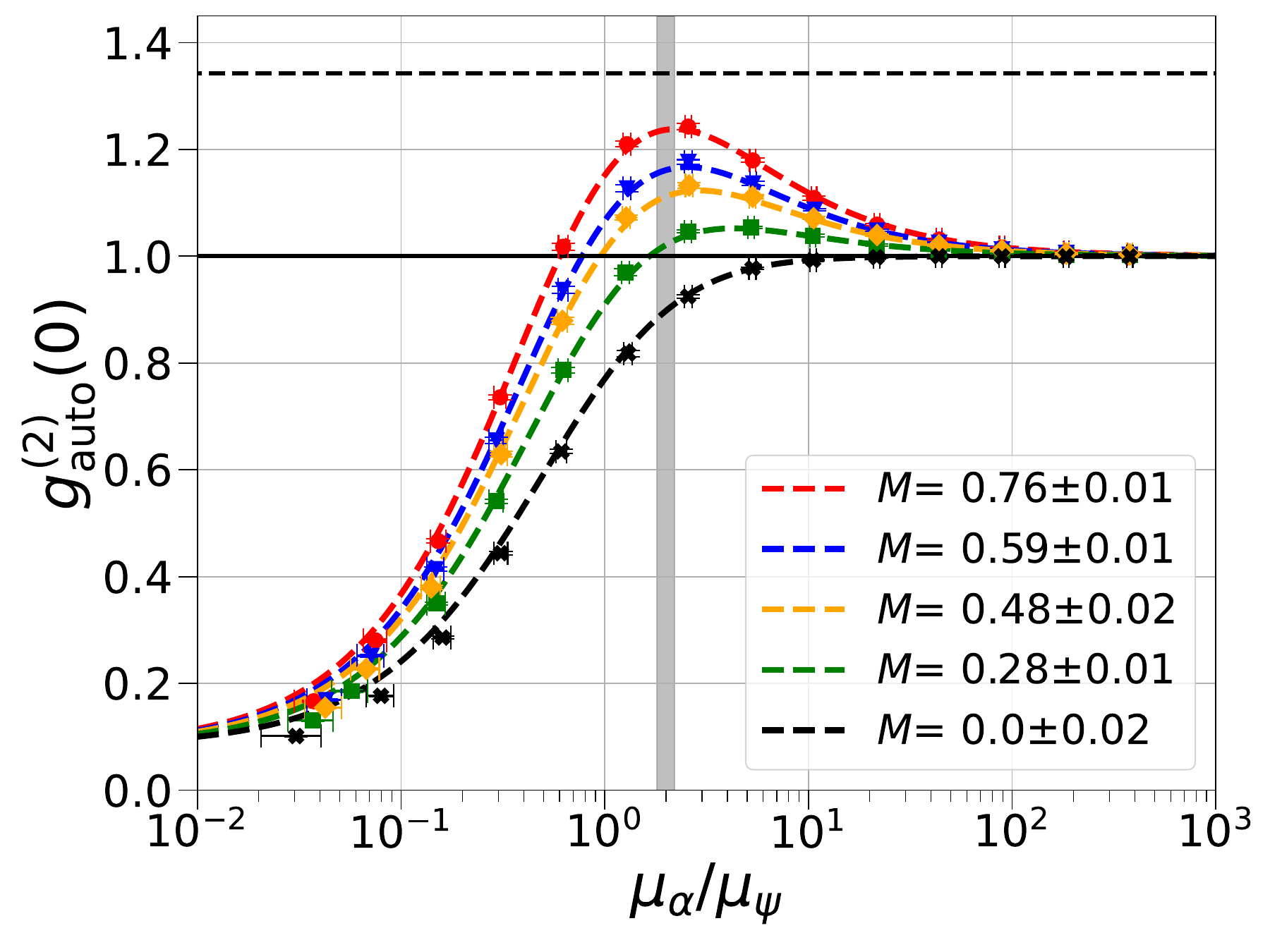}
    \caption{
   Photon correlations in one beam splitter output: $g^{(2)}_\mathrm{auto}(0)$ for various mean photon numbers $\mu_\alpha$. The mean wavepacket overlap $M_{}$ (different color) is controlled by adjusting the polarization of the mixing coherent state. In agreement with our theory (fit with Eq.(\ref{eq:g2_auto}), lines), with increasing power $\mu_\alpha$, the state in one FBS output shows a gradual transition from antibunching, through bunching, to random photon statistics.}\label{fig:g2_auto_plot}
\end{figure}

We now evaluate the mean wavepacket overlap between both fields by analyzing the bunching signature at one FBS output~\cite{Shen2017}. In Fig.~\ref{fig:g2_auto_plot}, we show different values of $g^{(2)}_\mathrm{auto}(0)$ as a function of the relative optical power of the fields $\mu_\alpha/\mu_\psi$. The values are derived from second-order correlation histograms $G^{(2)}_\mathrm{auto}(\tau)$ by normalizing with the uncorrelated peaks at $\tau\neq0$, shown in Fig.~\ref{fig:g2_cross_auto_sketch_histo}(c),

By controlling $\mu_\alpha$, we observe a gradual transition from the quantum regime of antibunching with $g^{(2)}_\mathrm{auto}(0)\approx 0$ in the single-photon stream at low $\mu_\alpha$ to classical coherent state statistics at high $\mu_\alpha$ ($g^{(2)}_\mathrm{auto}(0)\approx 1$). Interestingly, in the regime $\mu_\alpha/\mu_\psi\approx2$, where photon bunching is the strongest, $g^{(2)}_\mathrm{auto}(0)$ becomes highly sensitive to $M_{}$. For a pure single-photon stream ($g^{(2)}_\psi(0)=0$) and ideal overlap ($M_{}=1$), this value approaches the maximum of $\nicefrac{4}{3}$ for an ideal displaced single-photon Fock state \cite{Laiho2012}.

Our model (see Supplementary Section \ref{sec:Theory_supp}) allows to derive the expected second-order auto-correlation:
\begin{equation}
  g_{\mathrm{auto}}^{(2)}(0)=\frac{ \mu_\alpha^2+ \mu_\psi^2 g^{(2)}_\psi(0) +2\mu_\alpha \mu_\psi(1+M_{})}{\left( \mu_\alpha+\mu_\psi\right)^2}\,.  \label{eq:g2_auto}
\end{equation}
Intuitively, we identify three contributions: (i) the contribution $\mu_\alpha^2$, arising purely from the coherent state when all photons are reflected, (ii) the second term $\mu_\psi^2 g^{(2)}_\psi(0)$, representing two-photon correlations in the single-photon stream when all photons are transmitted, and (iii) the final term, containing information about the mode overlap where photons from both sources interfere, forming displaced Fock states \cite{Lvovsky2002}.

In Fig.~\ref{fig:g2_auto_plot}, we fit $g_{\mathrm{auto}}^{(2)}(0)$ with Eq.~(\ref{eq:g2_auto}) and extract values of the mean wavepacket overlaps $M_{}$. As expected, the extracted overlaps are consistent with values obtained from homodyne cross-correlation measurements performed simultaneously.

A nice feature of the auto-correlation measurement lies in the monotonic dependence of $g^{(2)}_\mathrm{auto}(0)$ on $M$, shown in Eq.~\ref{eq:g2_auto}. This property enables real-time optimization of the mutual overlap of the fields. For a fixed relative power of the fields $\mu_\alpha/\mu_\psi$, increasing the mode overlap results in stronger photon bunching and thus a higher value of ${g^{(2)}_\mathrm{auto}(0)}$,  as shown in a recording in the Supplementary material. Concretely, we first set $\mu_\alpha/\mu_\psi\approx2$, where the photon bunching is maximal, and fine-tune the optical and mechanical elements of the setup while systematically monitoring the value of ${g_\mathrm{auto}^{(2)}(0)}$. With an overall efficiency of the single-photon source and the setup of $\eta\approx3\,\%$, a short integration time of $1\,$s suffices for real-time feedback. We reliably achieve a maximum value of $g^{(2)}_\mathrm{auto} = 1.243\pm0.002$ at $\mu_\alpha/\mu_\psi\approx2$, corresponding to $M_{}=0.76\pm0.01$.

Finally, we note that the bunching analysis can be used to estimate the QD brightness at the input of an arbitrary beam splitter with transmission $T$ and reflection $R$ due to the relation between the strength of a power-calibrated laser needed to achieve the $g_{\mathrm{auto}}^{(2)}(0)$ maximum. Concretely, the maximum of $g_{\mathrm{auto}}^{(2)}(0)$ is observed at $T\mu_\alpha=2R\mu_\psi$~\cite{Steindl2023}, enabling to use the quantum interference as a local probe of the brightness $\mu_\psi$.

\section{Conclusion and discussion}

In conclusion, we have implemented two homodyne photon correlation techniques to control and quantify the mean wavepacket overlap $M$ between laser light pulses and on-demand single photons emitted by the QD-micropillar device. The mode overlaps are determined through an unambiguous relationship between photon bunching and field overlap, which we obtain by measuring the second-order intensity correlations at the single output or between both outputs of the beam splitter. Both methods give consistent values and are fully accounted for by our photon correlation model, which includes the statistical properties of the single-photon source, including its brightness and purity.

Generally, in addition to application in the continuous variable framework, the interference between laser light and single photons is used in many quantum protocols and quantum state generation schemes. Especially due to the inherent robustness of the correlation measurements against losses, the homodyne photon correlation techniques are not constrained by the detection efficiency. Therefore, we can exploit these techniques to improve quantum communication protocols for qubit teleportation, for instance~\cite{Pittman2003, Stevenson2013}, or to calibrate distant quantum nodes in quantum networks~\cite{Iuliano2024}. Finally, the interference of laser light and single photons can be used to generate high-NOON states~\cite{zhang_scalable_2018}, which are important resources for super-resolving phase measurements in quantum metrology~\cite{afek_high-noon_2010}.

\section{Acknowledgements}
H.L. acknowledges Leonid Vidro for the discussion. This work was partially supported by the Paris Ile-de-France Région in the framework of DIM QUANTIP, the European Union’s Horizon CL4 program under the grant agreement 101135288 for EPIQUE project, the Plan France 2030 through the projects ANR22-PETQ-0011, ANR-22-PETQ-0006, and ANR-22-PETQ-0013, and a public grant overseen by the French National Research Agency as part of the ”Investissements d’Avenir” programme (Labex NanoSaclay, reference: ANR10LABX0035). We acknowledge funding from NWO/OCW (Quantum Software Consortium, Nos. 024.003.037,024.003.037/3368). This work was carried out within the C2N micro-nanotechnologies platforms and was partially supported by the RENATECH network and the General Council of Essonne.

\newpage

\section{Supplementary}

\subsection{Derivation of photon correlations} \label{sec:Theory_supp}
In the experiment, we use short resonant $\pi$-pulse excitation of our QD leading to nearly ideal coherent population inversion and emission of a photonic state $\psi=\sum_n^2\sqrt{p_n}|n\rangle$, with $p_1\approx1$ and $p_0,p_2\ll p_1$ with negligible photon-number coherence. Therefore, we approximate the state as 
\begin{equation}
\hat{\rho} = p_1|1\rangle\langle1|+p_2|2\rangle\langle2|\,,\label{eq:QDstate_indist}
\end{equation}
enabling us to capture the reduced single-photon purity due to QD re-excitation (i.e. $p_2\neq0$). This state is propagated through a lossy optical channel with total transmission $\eta$ to a lossless balanced beam splitter, where it is mixed with the coherent state $|\alpha\rangle$. The polarization of the coherent state is controlled by a waveplate rotation by angle $\theta$ with respect to its fast axis aligned along the polarization of the QD light, enabling us to calculate the mode overlap $M=\cos^2\theta$ and to account for mutual indistinguishability of the two fields. This mixing leads to the generation of different displaced Fock states entangled between the beam splitter outputs (labeled 2 and 3)~\cite{Windhager2011,Dastidar2022}.

\begin{equation}
\hat{\rho}_\mathrm{out} = \mathscr{D}_2 \left(\alpha\right)\mathscr{D}_3 \left(\alpha\right) \hat{\rho}_{23}
\mathscr{D}_2 ^\dagger\left(\alpha\right)\mathscr{D}_3^\dagger \left(\alpha\right)\,,\label{eq:DisplacedState}
\end{equation}

with 

\begin{equation}
\begin{split}
\hat{\rho}_\mathrm{23} =& \tilde{p}_0|0,0\rangle\langle 0,0| +  {\frac{\tilde{p}_1}{2}} \left(|0,1\rangle\langle 1,0| +|0,1\rangle\langle 0,1| +|1,0\rangle\langle 1,0| +|1,0\rangle\langle 0,1| \right) \\
&+{\frac{\tilde{p}_2}{4}} \left(|0,2\rangle\langle 2,0| + |2,0\rangle\langle 0,2| + 2 |1,1\rangle\langle 1,1| \right) \\
&+ {\frac{\tilde{p}_2}{2\sqrt{2}}} \left(|1,1\rangle\langle 0,2| +|1,1\rangle\langle 2,0| +|2,0\rangle\langle 1,1|+|0,2\rangle\langle 1,1| \right) \,.
\end{split}
\end{equation}

We use notation $ \mathscr{D}_x \left(\alpha\right)=\hat{D}_{x, \parallel}\left(\frac{\alpha_\parallel}{\sqrt{2}}\right)\hat{D}_{x,\perp}\left(\frac{\alpha_\perp}{\sqrt{2}}\right)$, where the displacement operator $\hat{D}_{x,p}(\alpha,p)=e^{(\alpha_p a^{\dagger}_{x,p}-\alpha_p^{*}a_{x,p})}$ is acting in the spatial mode $x$ with polarization $p$; $\alpha_\parallel=\alpha\cos(\theta)$ and $\alpha_\perp=\alpha\sin(\theta)$ and loss-degraded probabilities $\tilde{p_0}=(1-\eta)(p_1+p_2-\eta p_2)$, $\tilde{p}_1=\eta(p_1+2p_2-2\eta p_2)$, and $\tilde{p}_2=\eta^2p_2$.

The state $\hat{\rho}_\mathrm{out}$ enables us to directly compute the photon correlations between the two beam splitter outputs 
\begin{equation}
G_{M}^{(2)}(0)=\langle \hat{\rho}_\mathrm{out} \hat{a}^\dagger_2 \hat{a}_2\hat{a}^\dagger_3 \hat{a}_3\rangle= \mu_\alpha^2+ \mu_\psi^2 g^{(2)}_\psi(0) +2\mu_\alpha \mu_\psi(1+M)\,,
\end{equation}
where $\mu_\psi=\eta(p_1+2p_2)$ is the end-to-end brightness of the QD source with $g^{(2)}_\psi(0)=2p_2/(p_1+2p_2)^2$, $\mu_\alpha=|\alpha|^2$ is mean photon number of the coherent state. The corresponding Hong-Ou-Mandel visibility is determined from the ratio between $G_{\mathrm{M}}^{(2)}(0)$ where the quantum interference is maximal and $G_{\mathrm{0}}^{(2)}(0)$ where the single-photon and coherent states Hilbert spaces are orthogonal ($M=0$)

\begin{equation}
V_\mathrm{HOM}=\frac{G_{\mathrm{0}}^{(2)}(0)-G_{\mathrm{M}}^{(2)}(0)}{G_{\mathrm{0}}^{(2)}(0)}\,,
\end{equation}
enabling to infer the field overlap \cite{maillette2023experimental, Iuliano2024}
\begin{equation}
M=V_{\mathrm{HOM}}\left(1+\frac{\mu_\alpha}{{2\mu_\psi}}g_\alpha^{(2)}(0)+\frac{\mu_\psi}{2\mu_\alpha}g_\psi^{(2)}(0)\right)\, .
\end{equation}

To evaluate the correlations, we have used: (i) the cyclic properties of the trace, (ii) the commutation relations between ladder and displacement operators $\hat{D}^\dagger (\alpha) \hat{a}^\dagger \hat{D}(\alpha)=\hat{a}^\dagger +\alpha^* $ and $\hat{D}^\dagger (\alpha)\hat{a} \hat{D}(\alpha)=\hat{a} +\alpha $ \cite{Oliveira1990}, (iii) unitarity of the displacement operator, and (iv) the relation $\hat{n}=\hat{n}_{\parallel}+\hat{n}_{\perp}$ connecting the polarization modes to non-polarized detection.

To evaluate the photon correlations at the single output of the beam splitter that performs the mixing, referred to in the main text as auto-correlation, we first post-select the state $\hat{\rho}_\mathrm{out}$ to only single output:

\begin{equation}
\hat{\rho}_\mathrm{2} = \mathrm{Tr}_3(\hat{\rho}_\mathrm{out})=  \mathscr{D}_2 \left(\alpha\right)\left[{A}_0|0\rangle\langle 0| + {{A}_1}|1\rangle\langle 1|+{{A}_2}|2\rangle\langle 2|\right] \mathscr{D}_2^\dagger \left(\alpha\right)\,.
\end{equation}
The post-selected state is a mixture of coherent state with weight $A_{0}=\tilde{p}_{0}+\frac{\tilde{p}_{1}}{2}+\frac{\tilde{p}_{2}}{4}$, displaced single-photon state with probability $A_{1}=\frac{1}{2}(\tilde{p}_{1}+\tilde{p}_{2})$, and displaced two-photon state with $A_{2}=\frac{\tilde{p}_{2}}{4}$. 

Now, we use the same properties (i)-(iv) as earlier, we evaluate the photon correlations at one output of the beam splitter
\begin{equation}
g_{\mathrm{auto}}^{(2)}(0)=\frac{\langle \hat{\rho}_{2} \hat{a}^\dagger_2 \hat{a}^\dagger_2 \hat{a}_2\hat{a}_2\rangle}{\langle \hat{\rho}_{2} \hat{a}^\dagger_2 \hat{a}_2\rangle^2}=\frac{ \mu_\alpha^2+ \mu_\psi^2 g^{(2)}_\psi(0) +2\mu_\alpha \mu_\psi(1+M_{})}{\left( \mu_\alpha+\mu_\psi\right)^2}\,.
\end{equation}
Due to the system symmetry given by the use of the balanced beam splitter for interference, the photon correlations $g_{\mathrm{auto}}^{(2)}(0)$ are up to normalization identical to the ones measured between both beam splitter outputs $G_{M}^{(2)}(0)$. Importantly, the photon correlations show a global maximum, enabling the use of simple gradient descent optimization to maximize the overlap $M$, which we use for real-time overlap optimization.

\subsubsection{Effect of partial distinguishability of QD photons}
Until now, we assumed that the QD emits fully indistinguishable photons in a single mode. However, due to charge and spin fluctuations in the QD environment, QD emission is often subject to small spectral wandering, which limits the indistinguishability of the consecutively emitted photons. We model this by mixing two density matrices corresponding to QD emission at the wavelength of the coherent state (subscript 0) and detuned with respect to this wavelength (subscript $\Delta$)
\begin{equation}
\hat{\rho}_{M_\psi} = M_\psi \hat{\rho} _0 +(1-M_\psi)\hat{\rho} _\Delta\,,\label{eq:QDstate_dist}
\end{equation}
where $M_\psi$ represents the QD emission indistinguishability. We mix $\hat{\rho}_{M_\psi}$ with the coherent state $|\alpha\rangle$ on the beam splitter, as before with $\hat{\rho}$. 
Here, only $\hat{\rho} _0$ would interfere with $|\alpha\rangle$, effectively reducing the measured overlap to $\tilde{M}=MM_\psi $ which propagates into the correlations functions
\begin{equation}
G_{\tilde{M}_{\mathrm{cross}}}^{(2)}(0)= \mu_\alpha^2+ \mu_\psi^2 g^{(2)}_\psi(0) +2\mu_\alpha \mu_\psi(1+\tilde{M}_{\mathrm{cross}})\,,
\end{equation}
and 
\begin{equation}
g_{\mathrm{auto}}^{(2)}(0)=\frac{ \mu_\alpha^2+ \mu_\psi^2 g^{(2)}_\psi(0) +2\mu_\alpha \mu_\psi(1+\tilde{M}_{\mathrm{auto}})}{\left( \mu_\alpha+\mu_\psi\right)^2}\,.
\end{equation}

\subsection{Coherent state mean photon-number calibration}\label{sec:Calib_mualpha}
We implement a scheme to control and estimate the mean photon number of the coherent state $\mu_\alpha$ at the point of interference by using the relation
\begin{equation}
    \mu_\alpha=P_{\!\alpha}\frac{\lambda\tau_\mathrm{rep}}{hc}
\end{equation}
with the Planck constant $h$, the speed of light $c$ and the power $P_{\!\alpha}$ at the point of interference. Two variable optical attenuators with total attenuation up to $\sim\!80\,$dB are placed before the free-space part of the coherent state to control the power, and a $(90:10)$ -FBS is added to constantly read out the power $P_0$ at the $90\,\%$ output using a power meter. The residual $10\,\%$ is sent to the mixing beam splitter. We calibrate the power meter to estimate the power after this beam splitter, i.e. the mean photon number $R\mu_\alpha=0.5 \mu_\alpha$, by connecting the respective FBS output fiber to a second power meter, indicated by the blue symbol in Fig.~\ref{fig:g2_cross_auto_sketch_histo}(a). The calibration
\begin{equation}
    P_{\!\alpha}=10^{-\mathcal{C}/10\,\mathrm{dB}} P_0
\end{equation}
yields an attenuation value of $\mathcal{C}\approx50\,$dB.

\subsubsection{Compensation for polarization-dependent detection efficiency}\label{sec:Suppl_polarizationCor}

The photons in our experiment are detected with SNSPDs, which are known to have polarization-dependent detection efficiency. We align and fix the detection polarization to maximize the detection counts of the photon stream $\psi$. This efficiency dependency affects the correlation measurements when we modify the mode overlap by adjusting the polarization of the mixing coherent state. Here, since the photons have two different polarizations, the detectors effectively filter out part of the incoming laser photons and thus artificially increase the measured bunching. We correct for this slightly reduced detection efficiency by increasing the mean photon number of the coherent state accordingly to detect the equal count rate for each polarization setting. We incorporate this adjustment into error estimation.

\subsection{Estimation of the mean wavepacket overlap \textit{M}}\label{subsec:MeanWavePacketOverlap_independent}

\begin{figure}[htbp]
    \centering
    \begin{subfigure}[b]{0.47\textwidth}
    \includegraphics[width=\textwidth]{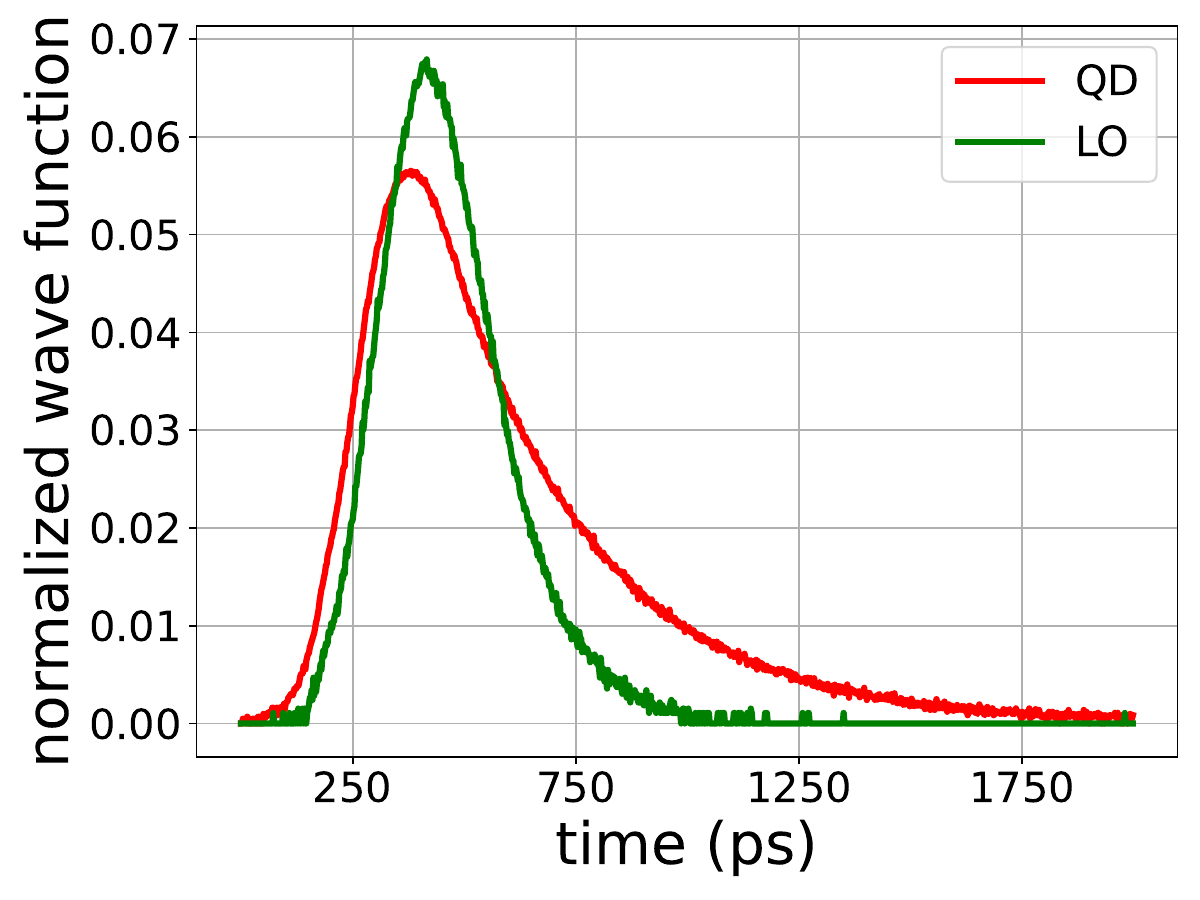}
    \label{fig:Overlap_TemporalMode}   
    \end{subfigure}\hfill
    \begin{subfigure}[b]{0.47\textwidth}
    \includegraphics[width=\textwidth]{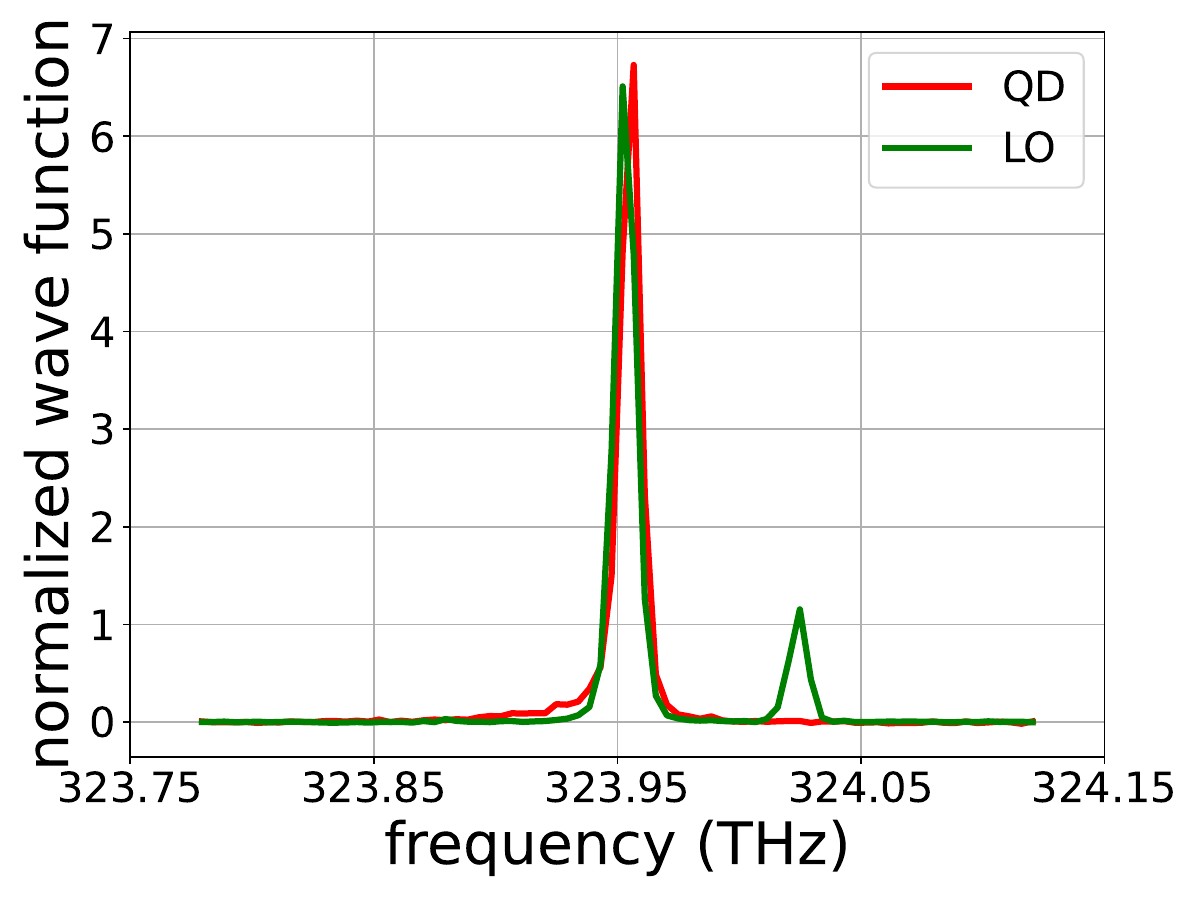}
    \label{fig:Overlap_FrequencyMode}            
    \end{subfigure}
    \caption{(a) Temporal and (b) frequency wavefunction overlap between light emitted by QD and the spectrally shaped laser.}
    \label{fig:Overlap_independent_TimeFreq}
\end{figure}
The mean wavepacket overlap $M=\Pi_x M_x$ between two fields can be estimated by a product of independently measured mode overlaps $M_x$ in individual degrees of freedom as
\begin{align}
   M_\mathrm{x} = \left| \int \frac{ \psi_1(x)}{\sqrt{\int  |\psi_1(\Tilde{x})|^2\,\mathrm{d}\Tilde{x}}}  \frac{ \psi_2(x)}{\sqrt{\int  |\psi_2(\Tilde{x})|^2\,\mathrm{d}\Tilde{x}}}\,\mathrm{d}x \right|^2, \label{eq:Overlap_wavefunction}
\end{align}
where $\psi_1(x)$ and $\psi_2(x)$ are the wavefunctions of the respective quantum states in this mode $x$.

First, we access the temporal overlap $M_t$, accounting for the arrival time and quantifying the similarity of the temporal profiles. We independently measured the temporal length of the wavepackets of $\psi$ and $\ket{\alpha}$ as the square root of the temporally resolved intensity signal and extracted $M_t=(91.0\pm0.5)\,\%$ -- normalized wave functions are shown in Fig.~\ref{fig:Overlap_independent_TimeFreq}(a). 

Second, we quantify with Eq.~\ref{eq:Overlap_wavefunction} the overlap in the frequency domain $M_f=(85\pm2)\,\%$ using the normalized spectra in Fig.~\ref{fig:Overlap_independent_TimeFreq}(b) measured with individual fields. Here, we clearly observe a second frequency peak, corresponding to off-resonant laser light leakage through the Fabry–Pérot etalons used for the laser light shaping. By additional spectral filtering around the QD emission to remove the off-resonant contribution done during post-processing, we retrieved $M_f=(91\pm2)\,\%$, limited by small relative detuning.

\begin{figure}[htbp]
    \centering
    \includegraphics[width=0.32\textwidth]{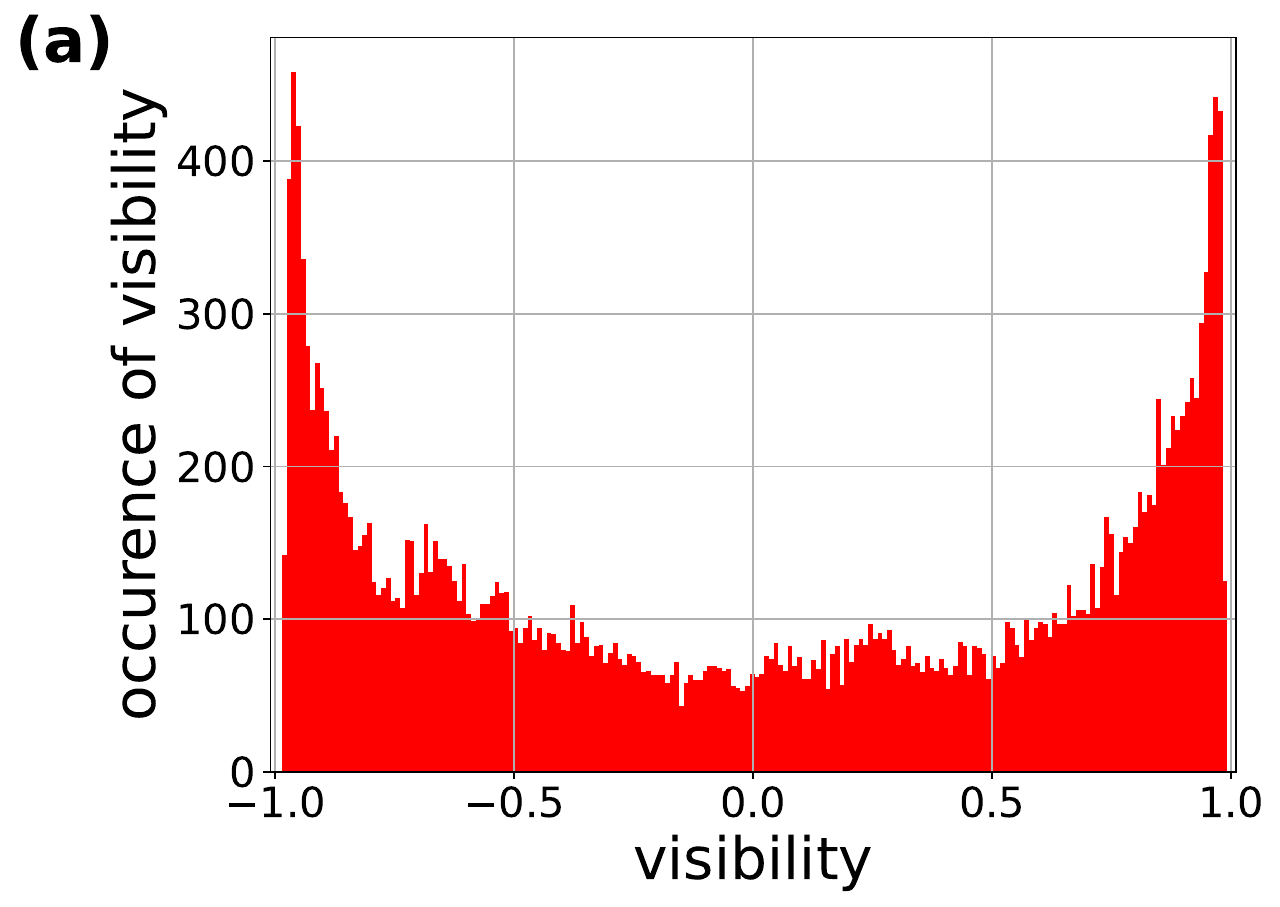}
    \includegraphics[width=0.32\textwidth]{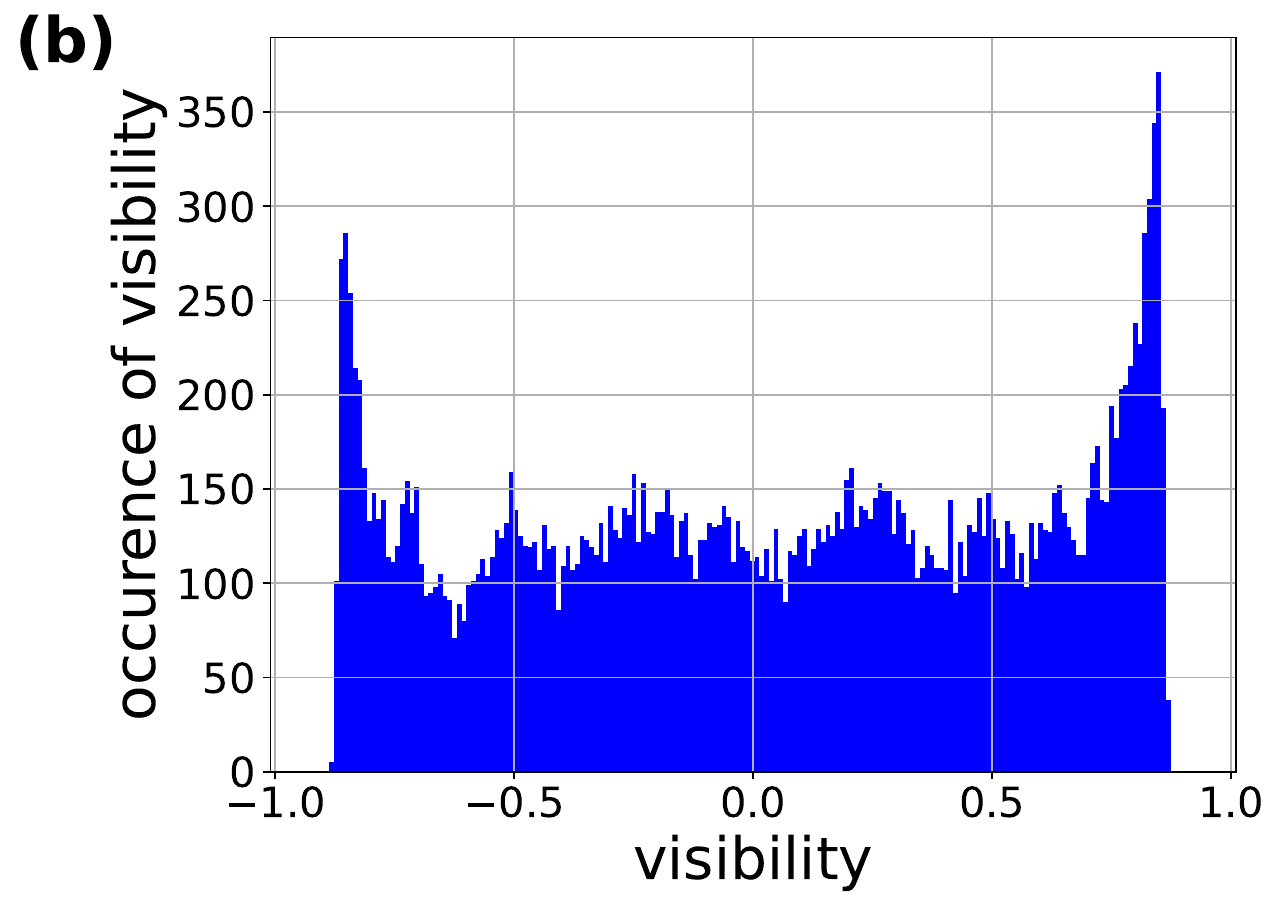}
    \includegraphics[width=0.32\textwidth]{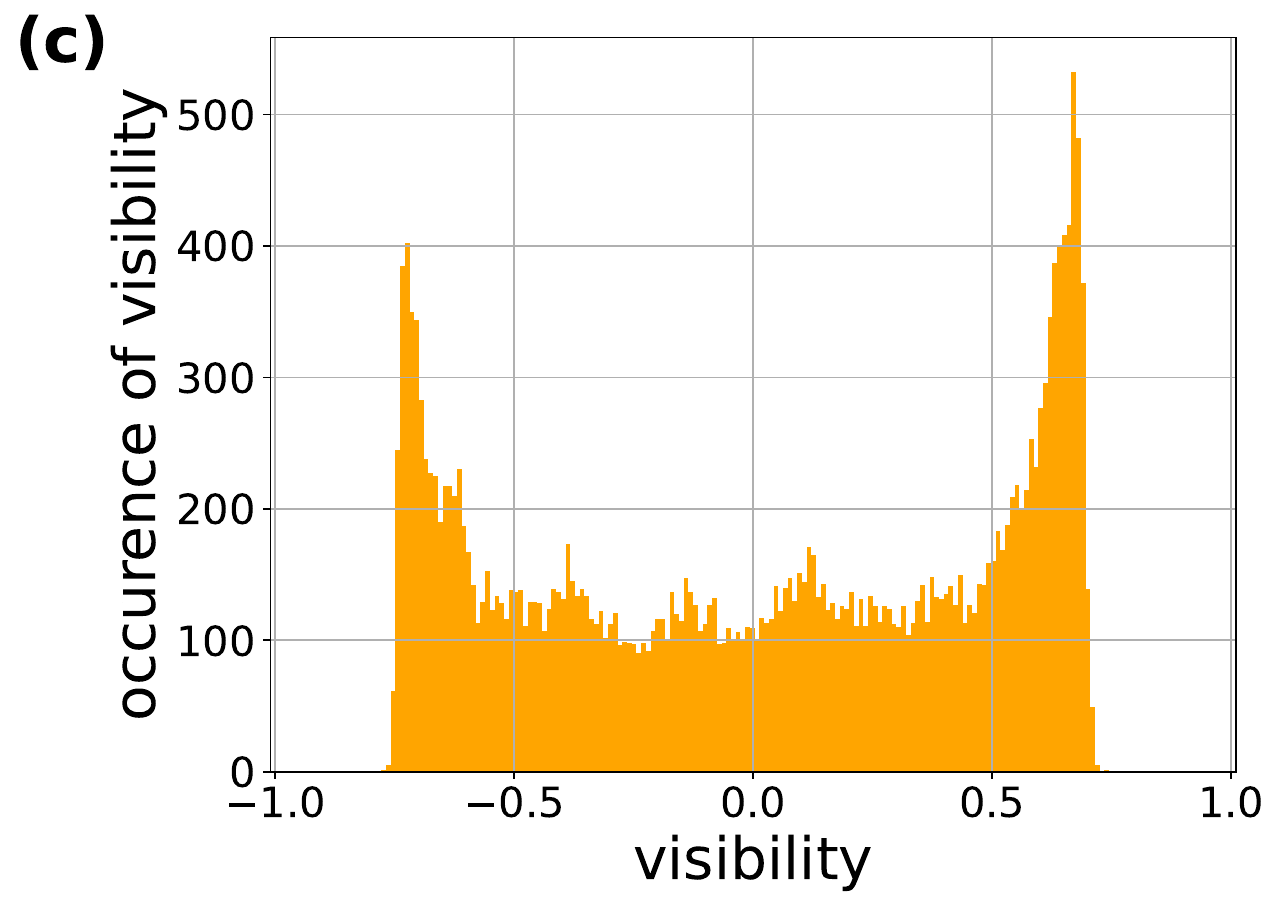}
    \includegraphics[width=0.32\textwidth]{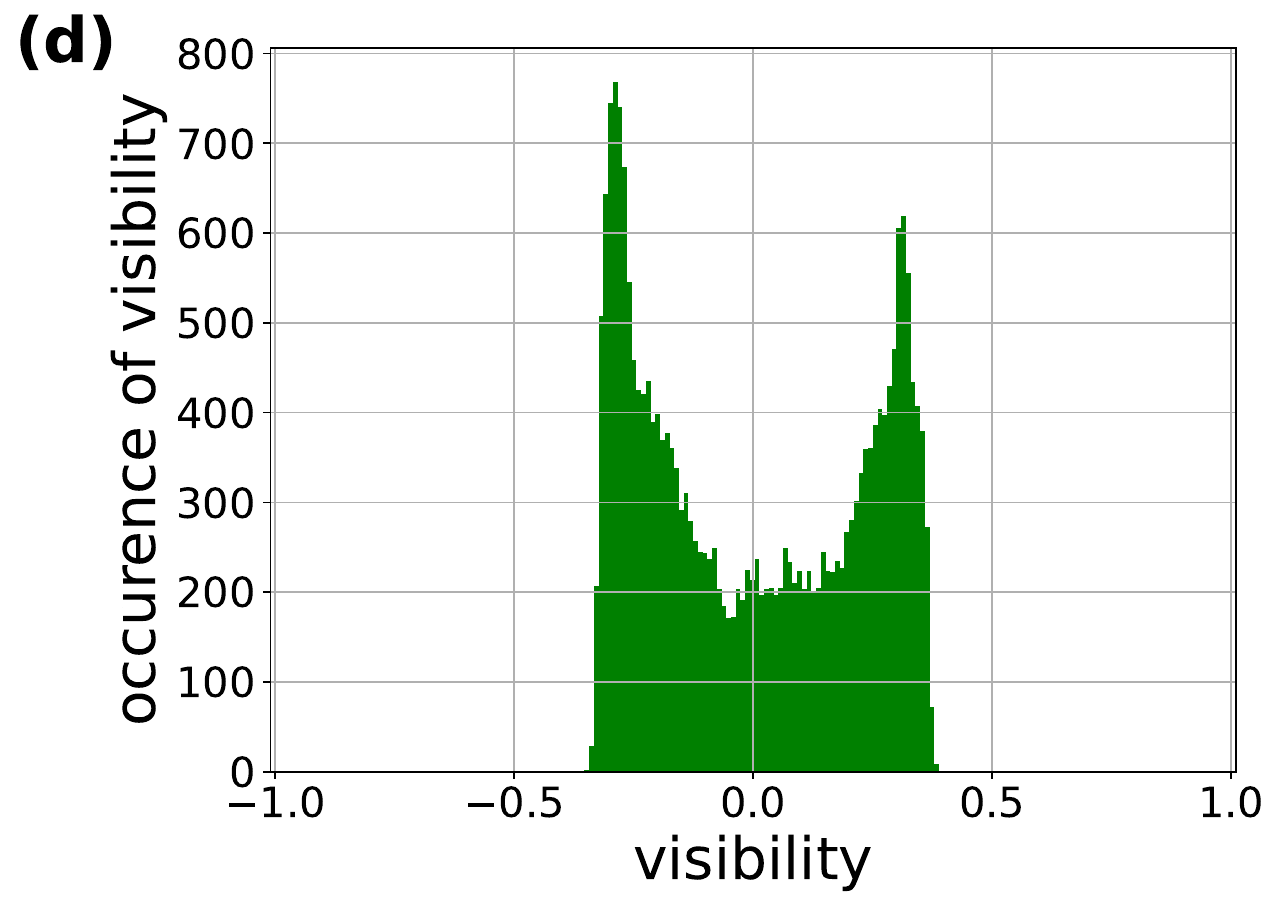}
    \includegraphics[width=0.32\textwidth]{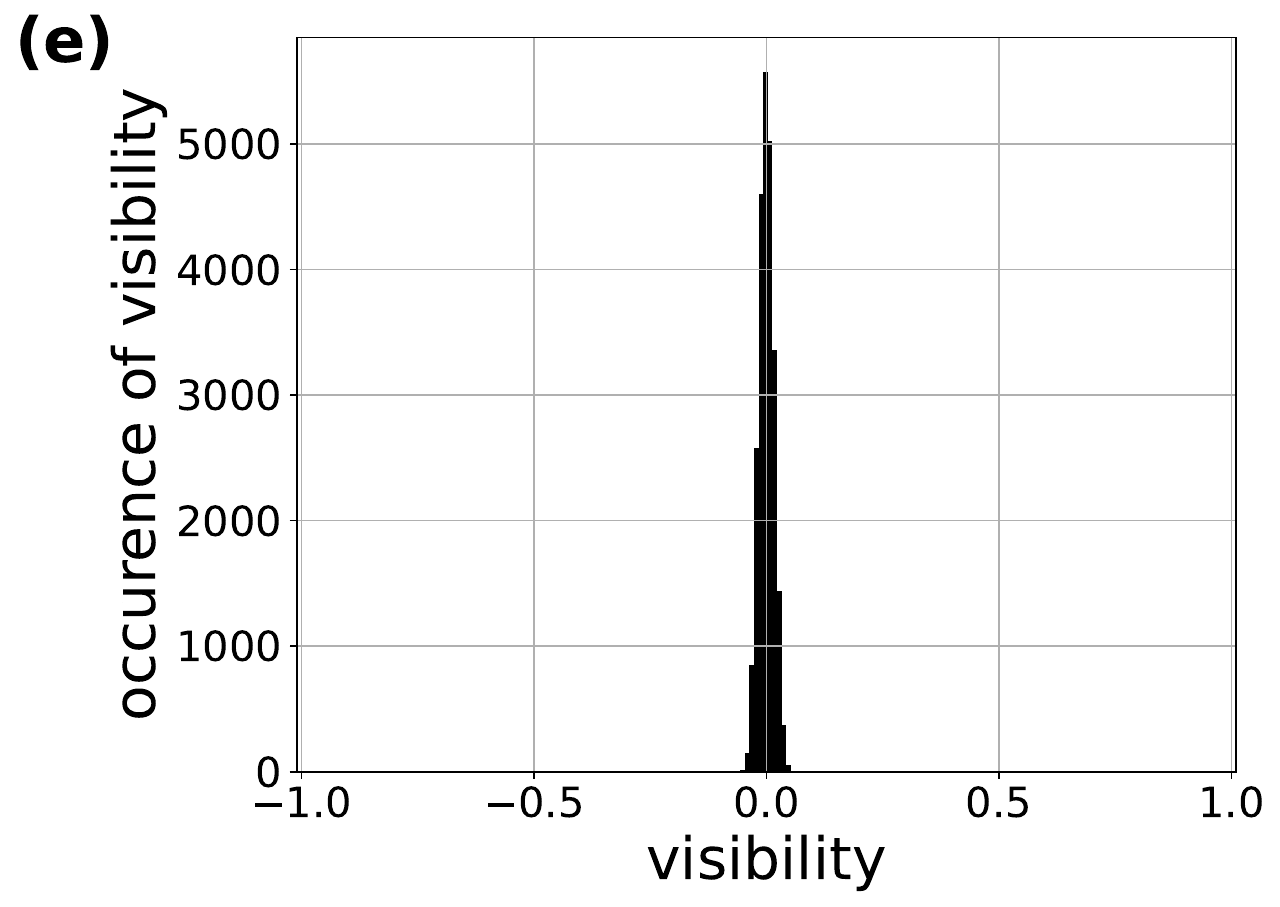}
    \includegraphics[width=0.32\textwidth]{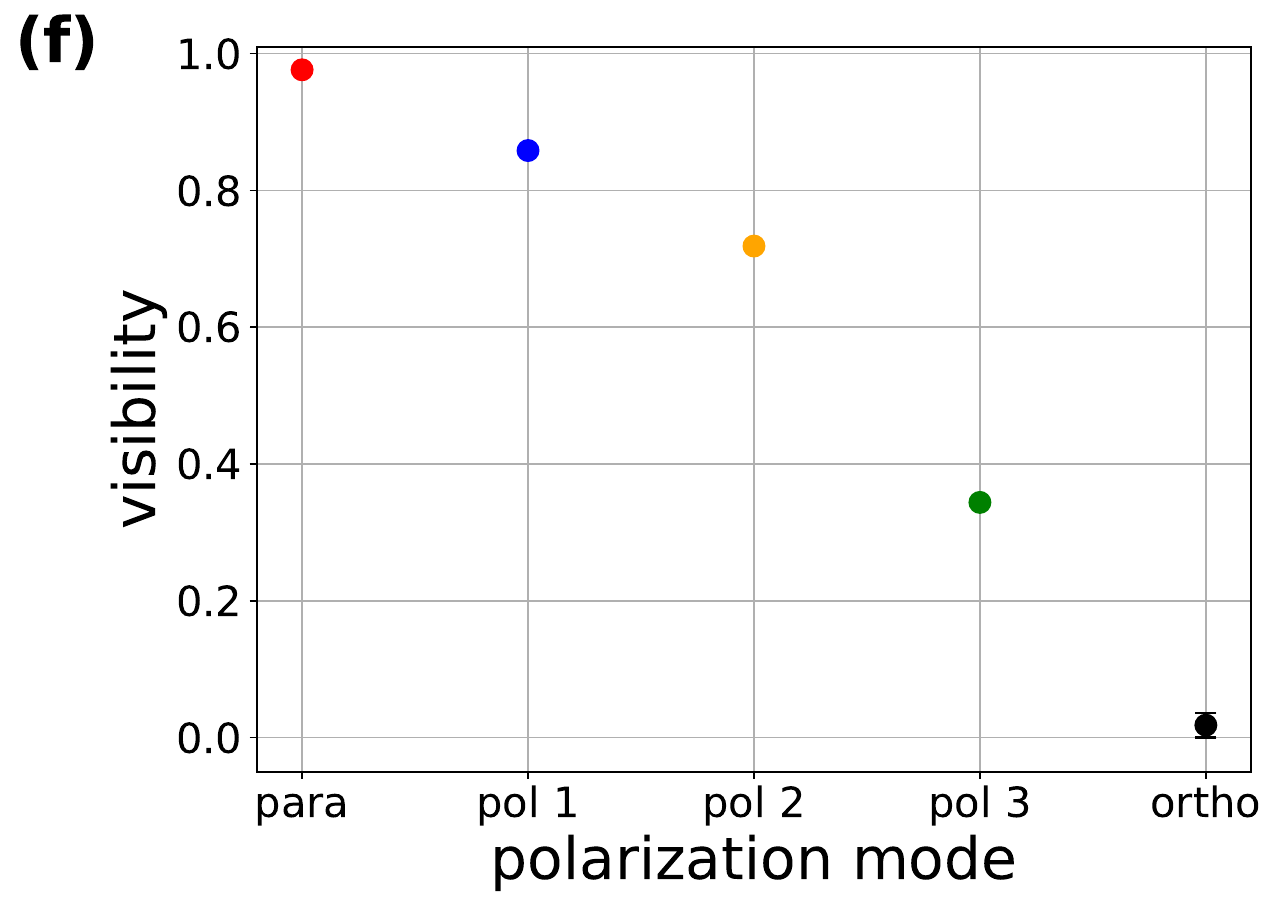}          
    \caption{(a-e) Visibility histograms measured for different settings of polarization. (f) The extracted visibility for the polarization settings shown in panels (a-e).}
    \label{fig:Overlap_Visibility}
\end{figure}

\subsection{Control of the overlap in the polarization degree of freedom \textit{M}}\label{subsec:polar}

\begin{table}[h!]
\centering
\begin{tabular}{lccccc}
        \toprule
         polarization & para $\parallel$ & pol 1&  pol 2 &  pol 3 & ortho $\perp$\\\midrule\midrule
         $M_tM_f$ $(\%)$ & $78\pm3$& $78\pm3$& $78\pm3$& $78\pm3$& $78\pm3$\\  
         $M_p$ $(\%)$ & $97.6\pm0.3$ & $86\pm2$& $72\pm2$& $34\pm2$ &  $0+2$ \\
         $M_tM_fM_p$ $(\%)$& $76\pm2$& $67\pm2$& $56\pm2$& $27\pm2$& $0.0+1.5$\\     \hline
         measured $M$ $(\%)$& $76\pm1$& $59\pm2$& $49\pm2$& $30\pm2$& $0+2$    \\
         \bottomrule
\end{tabular}
    \caption{Summary of the extracted mean wavepacket overlaps in individual modes $M_x$, independent estimation of $M$, and comparison to the measured overlap $M$.}
    \label{tab:Overlap_EstimationConclusion}
\end{table}
 
 The overlap in the polarization degree of freedom $M_p$ is determined from the two-detector visibility of the interference fringes measured at the output of the mixing beam splitter for different settings of relative polarization. To guarantee perfect mode overlap in the temporal and spectral domain, we interfere two beams derived from a single narrow-line continuous-wave (CW) laser. In Fig.~\ref{fig:Overlap_Visibility}(a), we show the visibility histogram measured with relative polarization aligned to achieve maximum visibility ($\parallel$, red). This visibility can be reduced by adjusting the relative polarization between the two interference paths; see Fig.~\ref{fig:Overlap_Visibility}(b-e) (pol 1, blue; pol 2, orange; pol 3, green; $\perp$, black). In the main text, we show photon-correlation measurements acquired for different relative polarizations, corresponding to the visibility histograms shown in Fig.~\ref{fig:Overlap_Visibility}(b-e) (pol 1, blue; pol 2, orange; pol 3, green; $\perp$, black). The mode overlap $M_p$ is calculated from each of the histograms by averaging the 500 highest and lowest visibility values, where the statistical variation is included in the error estimation. The reconstructed $M_p$ are summarized in Fig.~\ref{fig:Overlap_Visibility}(f). 

The total overlap for different settings of the polarization is then calculated $M=M_tM_fM_pM_s$ and summarized in Tab.~\ref{tab:Overlap_EstimationConclusion}, where we assume the spatial mode overlap $M_s=1$ guaranteed by the single-mode FBS.

\bibliography{main}

\begin{thebibliography}{10}
\newcommand{\enquote}[1]{``#1''}

\bibitem{maring2024versatile}
N.~Maring, A.~Fyrillas, M.~Pont, \emph{et~al.}, \enquote{A versatile single-photon-based quantum computing platform,} {\protect\JournalTitle{Nature Photonics}} pp. 1--7 (2024).

\bibitem{aghaee_rad_scaling_2025}
H.~Aghaee~Rad, T.~Ainsworth, R.~N. Alexander, \emph{et~al.}, \enquote{Scaling and networking a modular photonic quantum computer,} {\protect\JournalTitle{Nature}}  (2025).

\bibitem{chen_integrated_2021}
Y.-A. Chen, Q.~Zhang, T.-Y. Chen, \emph{et~al.}, \enquote{An integrated space-to-ground quantum communication network over 4,600 kilometres,} {\protect\JournalTitle{Nature}} \textbf{589}, 214--219 (2021). Publisher: Nature Publishing Group.

\bibitem{chen_twin-field_2021}
J.-P. Chen, C.~Zhang, Y.~Liu, \emph{et~al.}, \enquote{Twin-field quantum key distribution over a 511 km optical fibre linking two distant metropolitan areas,} {\protect\JournalTitle{Nature Photonics}} \textbf{15}, 570--575 (2021). Publisher: Nature Publishing Group.

\bibitem{hajomer_long-distance_2024}
A.~A.~E. Hajomer, I.~Derkach, N.~Jain, \emph{et~al.}, \enquote{Long-distance continuous-variable quantum key distribution over 100-km fiber with local local oscillator,} {\protect\JournalTitle{Science Advances}} \textbf{10}, eadi9474 (2024). Publisher: American Association for the Advancement of Science.

\bibitem{jia_squeezing_2024}
W.~Jia, V.~Xu, K.~Kuns, \emph{et~al.}, \enquote{Squeezing the quantum noise of a gravitational-wave detector below the standard quantum limit,} {\protect\JournalTitle{Science}} \textbf{385}, 1318--1321 (2024). Publisher: American Association for the Advancement of Science.

\bibitem{Lloyd1999}
S.~Lloyd and S.~L. Braunstein, \enquote{{Quantum computation over continuous variables},} {\protect\JournalTitle{Phys. Rev. Lett.}} \textbf{82}, 1784--1787 (1999).

\bibitem{Dakna1998}
M.~Dakna, L.~Kn{\"{o}}ll, and D.~G. Welsch, \enquote{{Quantum state engineering using conditional measurement on a beam splitter},} {\protect\JournalTitle{Eur. Phys. J. D}} \textbf{3}, 295--308 (1998).

\bibitem{Dakna1997}
M.~Dakna, T.~Anhut, T.~Opatrn{\'{y}}, \emph{et~al.}, \enquote{{Generating Schr{\"{o}}dinger-cat-like states by means of conditional measurements on a beam splitter},} {\protect\JournalTitle{Phys. Rev. A - At. Mol. Opt. Phys.}} \textbf{55}, 3184--3194 (1997).

\bibitem{Barnett2018}
S.~M. Barnett, G.~Ferenczi, C.~R. Gilson, and F.~C. Speirits, \enquote{Statistics of photon-subtracted and photon-added states,} {\protect\JournalTitle{Phys. Rev. A}} \textbf{98}, 013809 (2018).

\bibitem{Zavatta2007}
A.~Zavatta, V.~Parigi, and M.~Bellini, \enquote{{Experimental nonclassicality of single-photon-added thermal light states},} {\protect\JournalTitle{Phys. Rev. A}} \textbf{75}, 052106 (2007).

\bibitem{Lvovsky2002_photoncatalysis}
A.~I. Lvovsky and J.~Mlynek, \enquote{{Quantum-Optical Catalysis: Generating Nonclassical States of Light by Means of Linear Optics},} {\protect\JournalTitle{Phys. Rev. Lett.}} \textbf{88}, 250401 (2002).

\bibitem{Lvovsky2002}
A.~I. Lvovsky and S.~A. Babichev, \enquote{{Synthesis and tomographic characterization of the displaced Fock state of light},} {\protect\JournalTitle{Phys. Rev. A}} \textbf{66}, 011801 (2002).

\bibitem{eaton_non-gaussian_2019}
M.~Eaton, R.~Nehra, and O.~Pfister, \enquote{Non-{Gaussian} and {Gottesman}–{Kitaev}–{Preskill} state preparation by photon catalysis,} {\protect\JournalTitle{New Journal of Physics}} \textbf{21}, 113034 (2019).

\bibitem{Paris1996}
M.~G. Paris, \enquote{{Displacement operator by beam splitter},} {\protect\JournalTitle{Phys. Lett. A}} \textbf{217}, 78--80 (1996).

\bibitem{lvovsky_continuous-variable_2009}
A.~I. Lvovsky and M.~G. Raymer, \enquote{Continuous-variable optical quantum-state tomography,} {\protect\JournalTitle{Reviews of Modern Physics}} \textbf{81}, 299--332 (2009).

\bibitem{laiho2009producing}
K.~Laiho, K.~Cassemiro, and C.~Silberhorn, \enquote{Producing high fidelity single photons with optimal brightness via waveguided parametric down-conversion,} {\protect\JournalTitle{Optics express}} \textbf{17}, 22823--22837 (2009).

\bibitem{Koashi1996}
M.~Koashi, M.~Matsuoka, and T.~Hirano, \enquote{{Photon antibunching by destructive two-photon interference},} {\protect\JournalTitle{Physical Review A - Atomic, Molecular, and Optical Physics}} \textbf{53}, 3621--3624 (1996).

\bibitem{Laiho2012}
K.~Laiho, M.~Avenhaus, and C.~Silberhorn, \enquote{{Characteristics of displaced single photons attained via higher order factorial moments},} {\protect\JournalTitle{New J. Phys.}} \textbf{14}, 105011 (2012).

\bibitem{Shen2017}
A.~Shen, B.~Du, Z.-H. Wang, \emph{et~al.}, \enquote{Indistinguishability-induced classical-to-nonclassical transition of photon statistics,} {\protect\JournalTitle{Phys. Rev. A}} \textbf{95}, 053851 (2017).

\bibitem{Somaschi2016}
N.~Somaschi \emph{et~al.}, \enquote{Near-optimal single-photon sources in the solid state,} {\protect\JournalTitle{Nature Photonics}} \textbf{10}, 340--345 (2016).

\bibitem{Tomm2021}
N.~Tomm, A.~Javadi, N.~O. Antoniadis, \emph{et~al.}, \enquote{{A bright and fast source of coherent single photons},} {\protect\JournalTitle{Nat. Nanotechnol.}} \textbf{16}, 399--403 (2021).

\bibitem{Thomas2021}
S.~E. Thomas, M.~Billard, N.~Coste, \emph{et~al.}, \enquote{Bright polarized single-photon source based on a linear dipole,} {\protect\JournalTitle{Phys. Rev. Lett.}} \textbf{126}, 233601 (2021).

\bibitem{Ding2016}
X.~Ding, Y.~He, Z.-C. Duan, \emph{et~al.}, \enquote{On-demand single photons with high extraction efficiency and near-unity indistinguishability from a resonantly driven quantum dot in a micropillar,} {\protect\JournalTitle{Phys. Rev. Lett.}} \textbf{116}, 020401 (2016).

\bibitem{ding_high-efficiency_2025}
X.~Ding, Y.-P. Guo, M.-C. Xu, \emph{et~al.}, \enquote{High-efficiency single-photon source above the loss-tolerant threshold for efficient linear optical quantum computing,} {\protect\JournalTitle{Nature Photonics}}  (2025).

\bibitem{Vogel1995}
W.~Vogel, \enquote{Homodyne correlation measurements with weak local oscillators,} {\protect\JournalTitle{Phys. Rev. A}} \textbf{51}, 4160--4171 (1995).

\bibitem{Hong1987}
C.~K. Hong, Z.~Y. Ou, and L.~Mandel, \enquote{{Measurement of subpicosecond time intervals between two photons by interference},} {\protect\JournalTitle{Physical Review Letters}} \textbf{59}, 2044--2046 (1987).

\bibitem{ollivier2021hong}
H.~Ollivier, S.~Thomas, S.~Wein, \emph{et~al.}, \enquote{Hong-ou-mandel interference with imperfect single photon sources,} {\protect\JournalTitle{Physical Review Letters}} \textbf{126}, 063602 (2021).

\bibitem{Dousse2008}
A.~Dousse, L.~Lanco, J.~Suffczy\ifmmode~\acute{n}\else \'{n}\fi{}ski, \emph{et~al.}, \enquote{Controlled light-matter coupling for a single quantum dot embedded in a pillar microcavity using far-field optical lithography,} {\protect\JournalTitle{Phys. Rev. Lett.}} \textbf{101}, 267404 (2008).

\bibitem{Steindl2023_PER}
P.~Steindl, J.~Frey, J.~Norman, \emph{et~al.}, \enquote{Cross-polarization-extinction enhancement and spin-orbit coupling of light for quantum-dot cavity quantum electrodynamics spectroscopy,} {\protect\JournalTitle{Phys. Rev. Appl.}} \textbf{19}, 064082 (2023).

\bibitem{maillette2023experimental}
I.~Maillette~de Buy~Wenniger, S.~Thomas, M.~Maffei, \emph{et~al.}, \enquote{Experimental analysis of energy transfers between a quantum emitter and light fields,} {\protect\JournalTitle{Physical Review Letters}} \textbf{131}, 260401 (2023).

\bibitem{bennett2009interference}
A.~Bennett, R.~Patel, C.~Nicoll, \emph{et~al.}, \enquote{Interference of dissimilar photon sources,} {\protect\JournalTitle{Nature Physics}} \textbf{5}, 715--717 (2009).

\bibitem{Li2013}
L.~Li, Y.~O. Dudin, and A.~Kuzmich, \enquote{Entanglement between light and an optical atomic excitation,} {\protect\JournalTitle{Nature}} \textbf{498}, 466–469 (2013).

\bibitem{PadronBrito2021}
A.~Padr\'on-Brito, J.~Lowinski, P.~Farrera, \emph{et~al.}, \enquote{Probing the indistinguishability of single photons generated by rydberg atomic ensembles,} {\protect\JournalTitle{Phys. Rev. Res.}} \textbf{3}, 033287 (2021).

\bibitem{loredo_generation_2019}
J.~C. Loredo, C.~Antón, B.~Reznychenko, \emph{et~al.}, \enquote{Generation of non-classical light in a photon-number superposition,} {\protect\JournalTitle{Nature Photonics}} \textbf{13}, 803--808 (2019).

\bibitem{Prtljaga2016}
N.~Prtljaga, C.~Bentham, J.~O'Hara, \emph{et~al.}, \enquote{On-chip interference of single photons from an embedded quantum dot and an external laser,} {\protect\JournalTitle{Applied Physics Letters}} \textbf{108}, 251101 (2016).

\bibitem{Felle2015}
M.~Felle, J.~Huwer, R.~M. Stevenson, \emph{et~al.}, \enquote{Interference with a quantum dot single-photon source and a laser at telecom wavelength,} {\protect\JournalTitle{Applied Physics Letters}} \textbf{107} (2015).

\bibitem{Steindl2023}
P.~Steindl, \enquote{Quantum dots in microcavities: From single spins to engineered states of light,} Ph.D. thesis, Leiden University (2023).

\bibitem{Pittman2003}
T.~B. Pittman and J.~D. Franson, \enquote{Violation of bell's inequality with photons from independent sources,} {\protect\JournalTitle{Phys. Rev. Lett.}} \textbf{90}, 240401 (2003).

\bibitem{Stevenson2013}
R.~M. Stevenson, J.~Nilsson, A.~J. Bennett, \emph{et~al.}, \enquote{{Quantum teleportation of laser-generated photons with an entangled-light-emitting diode},} {\protect\JournalTitle{Nature Communications}} \textbf{4}, 2859 (2013).

\bibitem{Iuliano2024}
M.~Iuliano, M.-C. Slater, A.~J. Stolk, \emph{et~al.}, \enquote{Qubit teleportation between a memory-compatible photonic time-bin qubit and a solid-state quantum network node,} {\protect\JournalTitle{npj Quantum Information}} \textbf{10}, 107 (2024).

\bibitem{zhang_scalable_2018}
L.~Zhang and K.~W.~C. Chan, \enquote{Scalable {Generation} of {Multi}-mode {NOON} {States} for {Quantum} {Multiple}-phase {Estimation},} {\protect\JournalTitle{Scientific Reports}} \textbf{8}, 11440 (2018). Publisher: Nature Publishing Group.

\bibitem{afek_high-noon_2010}
I.~Afek, O.~Ambar, and Y.~Silberberg, \enquote{High-{NOON} {States} by {Mixing} {Quantum} and {Classical} {Light},} {\protect\JournalTitle{Science}} \textbf{328}, 879--881 (2010).

\bibitem{Windhager2011}
A.~Windhager, M.~Suda, C.~Pacher, \emph{et~al.}, \enquote{{Quantum interference between a single-photon Fock state and a coherent state},} {\protect\JournalTitle{Opt. Commun.}} \textbf{284}, 1907--1912 (2011).

\bibitem{Dastidar2022}
M.~G. Dastidar and G.~Sarbicki, \enquote{{Detecting entanglement between modes of light},} {\protect\JournalTitle{Phys. Rev. A}} \textbf{105}, 062459 (2022).

\bibitem{Oliveira1990}
F.~Oliveira, M.~S. Kim, and L.~Knight, \enquote{{Properties of displaced number states},} {\protect\JournalTitle{Phys. Rev. A}} \textbf{41}, 2645 (1990).

\end{thebibliography}

\end{document}